\documentclass[superscriptaddress,secnumarabic,
amssymb,amsmath,nobibnotes,aps,prd,showpacs,nofootinbib]{revtex4}%
\usepackage{graphicx}
\usepackage{epsf}
\usepackage{bm}
\usepackage{amsmath}
\usepackage{amsfonts}
\usepackage{amssymb}
\usepackage{epstopdf}
\usepackage{natbib}
\usepackage{color}%
\usepackage{float}
\providecommand{\U}[1]{\protect\rule{.1in}{.1in}}
\newcommand{\be}{\begin{equation}}
\newcommand{\ee}{\end{equation}}

\newcommand{\mincir}{\raise
-3.truept\hbox{\rlap{\hbox{$\sim$}}\raise4.truept\hbox{$<$}\ }}
\newcommand{\magcir}{\raise
-3.truept\hbox{\rlap{\hbox{$\sim$}}\raise4.truept\hbox{$>$}\ }}

\begin{document}

%\author{Jaume de Haro$^{}$\footnote{E-mail: jaime.haro@upc.edu}
%and Jaume Amor\'os$^{}$\footnote{E-mail: jaume.amoros@upc.edu}
%}

%\affiliation{Departament de Matem\`atica Aplicada I, Universitat
%Polit\`ecnica de Catalunya, Diagonal 647, 08028 Barcelona, Spain}

\title{Reheating constraints in quintessential inflation}

\author{Jaume de Haro\footnote{E-mail: jaime.haro@upc.edu}}
\affiliation{Departament de Matem\`atiques, Universitat Polit\`ecnica de Catalunya, Colom 11, 08222 Terrassa, Spain}

%\author{Jaume Amor\'os\footnote{E-mail: jaume.amoros@upc.edu}}
%\affiliation{Departament de Matem\`atiques, Universitat Polit\`ecnica de Catalunya, Diagonal 647, 08028 Barcelona, Spain}

\author{Llibert Arest\'e Sal\'o\footnote{E-mail: llibert.areste@estudiant.upc.edu} }
\affiliation{Departament de Matem\`atiques, Universitat Polit\`ecnica de Catalunya, Diagonal 647, 08028 Barcelona, Spain}

%\maketitle

\thispagestyle{empty}

\begin{abstract}

We study the consequences of reheating in quintessential inflation. From simple inflationary quintessential models introduced in \cite{hap, hap2}, we show that when the reheating
is due to the production of heavy massive particles conformally coupled with gravity, a viable model which matches with the current observational data \cite{Ade, Planck, bicep2}
is only possible for reheating temperatures that range between $1$ GeV and $10^4$ GeV. On the other hand, when the universe reheats via the production of massless particles,
the viability of the model is only possible when those particles are nearly conformally coupled with gravity, leading to a reheating temperature between $1$ MeV and $10^4$ GeV.
\end{abstract}

\vspace{0.5cm}


%\keywords{Inflation, Deflation, Reheating, Cosmological constant.}

\maketitle

\section{ Introduction}

In two recent papers \cite{hap, hap2}  some  families of quintessential inflation models have been obtained, coming from very simple polynomial potentials which fit well with the current observational data provided by Planck's team \cite{Ade, Planck} and the BICEP/Keck-Planck Collaboration \cite{bicep2}. The models unify the early inflationary period with the current cosmic acceleration and contain a phase transition from the inflationary phase to a kination regime, which is essential to produce enough particles to thermalize the universe with a reheating temperature compatible with the bounds coming from nucleosynthesis.

\

In the present work we explore the consequences and constraints that one obtains from the relation that exists between the spectral index and the reheating temperature. This relation comes from the fact that the number of e-folds could be obtained in two completely different ways: by definition through an expression that only depends on the spectral index or using the whole history of the universe that leads to a function of the reheating temperature and the spectral index.
On the other hand, to obtain simple expressions of the reheating temperature one can consider either the production of heavy massive particles conformally coupled with gravity or massless particles nearly conformally coupled with gravity. 

\

Therefore, for a given reheating temperature between $1$ MeV and $10^9$ GeV, which is required in order to have a successful nucleosynthesis \cite{bounds}, one obtains the corresponding value of the spectral index and, thus, the value of the  ratio of tensor to scalar perturbations. Since this ratio must be less than $0.12$ \cite{bicep2}, this constrains the reheating temperature to range between $1$ MeV and $10^4$ GeV. 
Moreover, since in the case of creation of heavy particles the mass of these produced particles must be less than Planck's mass -in the opposite case these particles would become micro black holes-, the model only supports temperatures greater than $1$ GeV and less than $10^4$ GeV. Finally, dealing with massless particles, we have shown that the viability of the model is only possible when the coupling constant is very close to $1/6$, i.e., the created particles have to be nearly conformally coupled with gravity.

\

{

This low reheating temperature coming from the constraints that have been found in this paper leads to ensure a successful baryogenesis with thermal equilibrium \cite{davidson} ($1\text{ GeV }\lesssim T_R\ll 100$ GeV). Our model    also 
supports 
a unified  origin
of baryons and dark matter,
 via the
decays of a heavy scalar field which dominates the universe before nucleosynthesis \cite{kitano}, and avoids the
  cosmological moduli and gravitino  problems. }

\

The paper is organized as follows: In Section II, we depict all the characteristics of the simplest model of the family, which corresponds to a potential whose inflationary piece is a quadratic one and that, in order to have a phase transition to a kination regime, is matched with a small cosmological constant which is responsible for the current cosmic acceleration.
Section III is devoted to relate the reheating temperature, obtained from different particle production after the phase transition, with the spectral parameters,
which leads to a reheating temperature less than $10^4$ GeV in order to match with the current observation data. Finally, we have introduced an Appendix where we prove one of the basic formulas of this work: the energy density of the heavy massive particles conformally coupled with gravity. In \cite{hap, hap2} this quantity has been calculated using the concept of {\it number of particle produced in curved spacetimes}, which does not have a universal consensus \cite{particles}, whereas here we obtain this quantity directly from the definition of energy density and using the ``adiabatic prescription"  (see \cite{parker} for a review of the method) to regularize it.

\vskip 0.3cm

The units used throughout the paper are $\hbar=c=1$ and, with these units, $M_{pl}=\frac{1}{\sqrt{8\pi G}}$ is the reduced Planck's mass.

\section{Review of the model}
\label{model}

In a recent paper \cite{hap2} one of the authors of the present work and his collaborators introduced a family of models depending on a parameter $\alpha$,  which matches, at $2\sigma$ C.L.,  with the current observational data 
provided by Planck's team \cite{Ade,Planck}
when  $\alpha\in [0,\frac{1}{2}]$. Here, we consider the simplest one which corresponds to $\alpha=0$ and which describes a universe containing a small cosmological constant $\Lambda$ and filled with a fluid with a very simple linear Equation of State (EoS), which has a sudden phase transition when the energy density is $\rho=\rho_E$, 
\begin{eqnarray}\label{EoS}
 P=\left\{\begin{array}{ccc}
      -\rho+2{\rho_E}  & \mbox{for}&\rho\geq \rho_E\\
      \rho& \mbox{for}& \rho\leq \rho_E ,
             \end{array}\right.
\end{eqnarray}
where $\rho_E=(3H_E^2-\Lambda)M_{pl}^2$, being $H_E$ the parameter of the model. A simple calculation leads to the following dynamics
\begin{eqnarray}\label{dynamics}
\dot{H}=\left\{\begin{array}{ccc}
-3H_E^2+\Lambda& \mbox{for}& H\geq H_E\\
-3H^2+\Lambda& \mbox{for}& H\leq H_E,
\end{array}\right.
\end{eqnarray}
 which can be analytically solved leading to the following Hubble parameter 
\begin{eqnarray}
 H(t)=\left\{\begin{array}{ccc}
      \left(-3H_E^2+{\Lambda}  \right)t +H_E &\mbox{when}& t\leq 0\\
      \sqrt{\frac{\Lambda}{3}}\frac{3H_E+\sqrt{3\Lambda}\tanh(\sqrt{3\Lambda} t)}{3H_E\tanh(\sqrt{3\Lambda} t)+\sqrt{3\Lambda}} &\mbox{when}& t\geq 0,
             \end{array}\right.
\end{eqnarray}
and the corresponding scale factor is
\begin{eqnarray}
 a(t)=\left\{\begin{array}{ccc}
      a_E e^{\left[\left(-3H_E^2+{\Lambda}  \right)\frac{t^2}{2}+H_Et\right]}
       & \mbox{when}&t\leq 0\\
    a_E \left(\frac{3H_E}{\sqrt{3\Lambda}}\sinh(\sqrt{3\Lambda} t)+\cosh(\sqrt{3\Lambda} t)\right)^{\frac{1}{3}}  & \mbox{when}&t\geq 0.
             \end{array}\right.
\end{eqnarray}

Moreover,
the effective EoS parameter, namely $w_{eff}$, which is defined as $w_{eff}\equiv \frac{P}{\rho}=-1-\frac{2\dot{H}}{3H^2}$, for our  model is given by
\begin{eqnarray}
 w_{eff}=\left\{\begin{array}{ccc}
      -1+\frac{2}{3H^2}\left(3H_E^2-\Lambda\right)   &\mbox{when}& H\geq H_E\\
      1- \frac{2\Lambda}{3H^2}& \mbox{when}& H\leq H_E,
             \end{array}\right.
\end{eqnarray}
which shows that for $H\gg H_E$ one has
$w_{eff}(H)\cong -1$, meaning that we have an early inflationary quasi de Sitter period. As has been proved in \cite{hap2}, this model provides theoretical values of the spectral index and the ratio of tensor to scalar perturbations which enters, at $2\sigma$ C.L.,  in the $2$-dimensional marginalized joint confidence contour in the plane $(n_s,r)$, without the presence of running in the spectral index.  
At the phase transition, i.e. when $H\cong H_E$, the EoS parameter satisfies
$w_{eff}(H)\cong 1$ 
and the universe enters in a
kination or deflationary period \cite{spokoiny,joyce}, and finally, for $H\cong  \sqrt{\frac{\Lambda}{3}}$ one also has  $w_{eff}(H)\cong -1$
depicting the  current cosmic acceleration.

\

{Note that
the dynamics in \eqref{dynamics} could also be obtained assuming that the universe is filled with a viscous fluid. Effectively, in this case the Raychauduri equation becomes \cite{brevik13}
\begin{equation}
\dot{H}= -\frac{3}{2}(1+w)H^2+\frac{3\zeta}{2 M_{pl}^2}H+\frac{1}{2}(1+w)\Lambda. \label{1.31}
\end{equation}
Therefore, taking 
 $w=1$ and  the  following viscosity coefficient 
\begin{eqnarray}\zeta=\left\{\begin{array}{cc}
{2}M_{pl}^2\left(H-\frac{H_E^2}{H}\right)& H\geq H_E\\
0& H\leq H_E,
\end{array}\right.
\end{eqnarray}
one obtains the dynamics \eqref{dynamics} after  inserting it in \eqref{1.31}.
}

\

On the other hand,
using the reconstruction method \cite{hap2,hap}, one can also calculate 
the quadratic potential that leads to this dynamics
\begin{eqnarray}\label{h16}
V(\phi)=\left\{\begin{array}{ccc}
\frac{9}{2}\left(H_E^2-\frac{\Lambda}{3}\right)\left( \phi^2-\frac{2}{3}M_{pl}^2 \right)& \mbox{for} & \phi\leq \phi_E\\
{\Lambda}M_{pl}^2& \mbox{for} & \phi\geq \phi_E,
\end{array}\right.
\end{eqnarray}
where  the value of the scalar field at the transition time is $\phi_E\equiv -\sqrt{\frac{2}{3}}\frac{H_E M_{pl}}{\sqrt{H_E^2-\frac{\Lambda}{3}}}\cong -\sqrt{\frac{2}{3}}{ M_{pl}}$ and we have included the cosmological potential into
the potential.

\

Finally,
the number of e-folds from when the pivot scale leaves the Hubble radius until the end of inflation, and the main slow-roll parameter when the pivot scale crosses the Hubble radius, in terms of the spectral index, can be calculated exactly in this model, giving as a result
\begin{eqnarray}\label{A}
N=\frac{1}{2}\left(\frac{4}{1-n_s} -1  \right),\qquad     
\epsilon\equiv -\frac{\dot H}{H^2}=\frac{1-n_s}{4} \Longrightarrow  r=16\epsilon=4(1-n_s).
\end{eqnarray}
%where the stars means that the quantities are evaluated when the pivot scale crosses the Hubble radius.

We note also that the value of the parameter of the model $H_E$ is obtained from the theoretical \cite{btw} and the observational \cite{bld} value of the power spectrum
\begin{eqnarray}\label{power1}
 {\mathcal P}\cong \frac{H^2}{8\pi^2\epsilon M_{pl}^2}\sim 2\times 10^{-9},
\end{eqnarray}
using that  for our model $H\cong\frac{H_E}{\left(\frac{\epsilon}{3} \right)^{\frac{1}{2}}}$ and  $\epsilon=\frac{1-n_s}{4}$,  obtaining 
\begin{eqnarray}
 H_E \sim 7\times 10^{-4}\left(\frac{1-n_s}{12}\right) M_{pl}.
\end{eqnarray}

\section{Reheating constraints}

The model contains a phase transition from inflation to kination which is essential to create enough { $\chi$-particles} which reheat the universe \cite{spokoiny,pv}. When the produced particles are 
very massive and  conformally coupled with gravity satisfying $H_E\ll m\leq  M_{pl}$, its energy density evolves like (see the Appendix)
{\begin{eqnarray}
\rho_{\chi}\sim 4\times10^{-3} \frac{H_E^6}{m^2} \left( \frac{a_E}{a} \right)^3,
\label{rhomassive}
\end{eqnarray}}
where $a_E$ is the scale factor at the transition time. { Then, to thermalize the universe, we assume that these particles decay into lighter particles 
 in the same way as depicted in  \cite{pv} and \cite{37}, i.e., via  the exchange of gauge bosons whose thermalization rate  is given by $\Gamma=\sigma \frac{\rho_{\chi}}{m}$,
 where the cross section $\sigma$ for a $2\rightarrow 3$ scattering is $\sigma=\beta^3\rho^{1/2}_{\chi}$ with, as usual,  $\beta^2\sim 10^{-3}$ \cite{spokoiny}. In these conditions
 the reheating temperature  is of the order \cite{hap2}}
\begin{eqnarray}\label{temperature}
 T_R\sim 
 10^{-1}\left(\frac{H_E}{M_{pl}} \right)^2\left(\frac{H_E}{m}\right)M_{pl}\sim 3\times 10^{-11}
 \left(\frac{1-n_s}{12} \right)^3 \frac{M_{pl}}{m}M_{pl}.\end{eqnarray}
 
 {
Here it is important to stress that formula \eqref{temperature} is only valid for masses smaller than the reduced Planck mass, because for $m\gtrsim M_{pl}$ the produced particles become micro black holes. Effectively, an elementary
particle of mass $m$ has a Compton wavelength $1/m$ and the corresponding Schwarzschild radius
is  $\frac{m}{4\pi M_{pl}^2}$. Then, for $m\geq \sqrt{4\pi}M_{pl}$ the Compton wavelength
is smaller than its Schwarzschild radius \cite{fkl}.
For these micro black holes, Hawking's formulas about evaporation are not applicable because in this case the emitted particles have masses of the order of Planck's mass or greater \cite{Giddings}, which means that
the thermodynamic description breaks down at these scales and it is unknown whether or how they radiate \cite{Helfer}. 
}

 \

On the other hand,  when one only considers massless particles nearly conformally coupled with gravity,
 its energy density decays as  \cite{ford,pv}
 {\begin{eqnarray}
\rho_{\chi}\sim {\mathcal N}\left(\xi-\frac{1}{6} \right)^2 H_E^4\left( \frac{a_E}{a} \right)^4,
\end{eqnarray}}
where $\xi$ is the coupling constant
and ${\mathcal N}=\frac{1}{8\pi^2}\int_0^{\infty}s|g(s)|^2ds$ where $g(s)=\frac{1}{H_E^4a_E^4}\int_{-\infty}^{\infty}e^{-2is\tau}a^2(\tau)R(\tau)d\tau$. By integrating twice by parts it is straightforward to prove that $|g(s)|$ is of order $1/s^2$ for large values of $s$ because $R(\tau)$ is continuous at the transition time. Hence, ${\mathcal N}$ is not UV-divergent. Moreover, proceeding analogously as in \cite{hap}, we can compute an upper bound for ${\mathcal N}$ by using Cauchy-Schwarz inequality and Plancherel Theorem,

\begin{eqnarray}
{\mathcal{N}}\leq \frac{1}{32\pi H_E^4a_E^4}\sqrt{\int_{-\infty}^{\infty}a(t)\left(\frac{d(a^2(t)R(t))}{dt} \right)^2dt}\sqrt{\int_{-\infty}^{\infty}a^3(t)R^2(t)dt}\approx 3.59,
\end{eqnarray}
showing that the value ${\mathcal N}$ is convergent. Moreover, we have been able to compute numerically a lower bound, i.e., ${\mathcal N}\geq \frac{1}{8\pi^2} \int_0^{20} s|g(s)|^2ds\approx 2.12$. Therefore, we can state that ${\mathcal N}$ is of order 1. In this case the reheating temperature becomes 
 \begin{eqnarray}\label{temperature1}
 T_R\sim {\mathcal N}^{3/4} \left|\xi-\frac{1}{6} \right|^{3/2} \frac{H_E^2}{M^2_{pl}} M_{pl}\sim 5\times 10^{-7} \left|\xi-\frac{1}{6} \right|^{3/2}
 \left(\frac{1-n_s}{12} \right)^2 M_{pl}.
 \end{eqnarray}

On the other hand, 
the number of e-folds can also be calculated using the well-known formula \cite{rg}
\begin{eqnarray}
\frac{k}{a_0H_0}
=e^{-N}\frac{H}{H_0}\frac{a_{end}}{a_E}\frac{a_E}{a_R}\frac{a_R}{a_M}\frac{a_M}{a_0}
\end{eqnarray}
being $k$ the pivot scale and
where ``end'', $R$ , $M$ and $0$ respectively symbolize the end of inflation,  the beginning of radiation era, the beginning of matter domination era (when the energy density of radiation and matter is the same) and the value of any quantity at the current time. 

\

For our model one obtains
\begin{eqnarray}
N=70.94-\frac{1}{6}\left[2-
\ln\left(\frac{1-n_s}{12} \right)\right]-\frac{1}{3}\ln\left(\frac{g_R^{\frac{1}{4}}T_R}{\mbox{GeV}} \right),
\end{eqnarray}
{where  for the effective number of degrees of freedom  $g_R$ we use the data given in \cite{rg}:
 $g_R = 107, \ 90$ and  $11$ for $T_R \geq 175$  GeV, $175$  {GeV} $ \geq  T_R \geq   200$  {MeV} and $200$  {MeV}  $\geq T_R  \geq 1$ {MeV}, respectively.
 Of course, this number is not exactly constant in these intervals (see for instance the Figure 1 of \cite{Husdal}), but this does not affect the results  provided by our formula because the  effective number
 appears on it as  $\frac{1}{12}\ln g_R$.}

\

This quantity has to be equal to the number of e-folds calculated in formula \eqref{A}, obtaining a relation between the spectral index and the reheating temperature

\begin{eqnarray}\label{X}
\frac{1}{2}\left(\frac{4}{1-n_s} -1  \right)=70.94-\frac{1}{6}\left[2-
\ln\left(\frac{1-n_s}{12} \right)\right]-\frac{1}{3}\ln\left(\frac{g_R^{\frac{1}{4}}T_R}{\mbox{GeV}} \right);
\end{eqnarray}
hence, for a given temperature between $1$ MeV and $10^9$ GeV one obtains the value of the spectral index for our model. And once one has this value for a given temperature, the mass of the quantum field can be computed using formula \eqref{temperature} or the coupling constant with \eqref{temperature1} when one deals with massless particles.

\

Defining $Y=\frac{12}{1-n_s}$,  and $F(Y)= Y+\ln Y $ one obtains the equation

\begin{eqnarray}
Y+\ln Y= 426.64 -2\ln\left(\frac{g_R^{\frac{1}{4}}T_R}{\mbox{GeV}} \right).\end{eqnarray}
which, given $T_R$,  only has a solution, because the  function $Y+\ln Y$ is monotone.

\subsection{Heavy massive particles conformally coupled with gravity}

Then, for heavy massive particles, if one chooses 
\begin{enumerate}\item
$T_R=1$ MeV, one obtains 
\begin{eqnarray}
Y+\ln Y=439.25,
\end{eqnarray}
which leads to the following value of the spectral index $n_s\cong 0.9723$, and a ratio of tensor to scalar perturbations equal to $r=0.1108$, where we have used that
$r=4(1-n_s)$.

Further, using \eqref{temperature} one obtains the following mass $m\sim  10^3 M_{pl} $. The value of the Hubble parameter at the transition time will be
$H_E\sim 10^{-6} M_{pl}$, and its value when the pivot scale leaves de Hubble radius is $H\sim 2\times 10^{-5} M_{pl}$. Finally the number of e-folds
is $N=72$.

\item $T_R=1$ GeV, one finds
\begin{eqnarray}
Y+\ln Y=424.39,
\end{eqnarray}
which leads to the following value of the spectral index $n_s\cong 0.9713$, and a ratio of tensor to scalar perturbations equal to $r=0.1148$.

Further,  using \eqref{temperature} one obtains the following mass $m\sim  M_{pl}$. The value of the Hubble parameter at the transition time will be
$H_E\sim  10^{-6} M_{pl}$, and its value when the pivot scale leaves de Hubble radius is $H\sim   2\times 10^{-5} M_{pl}$. Finally the number of e-folds
is $N=69 $.

\item $T_R=10^4$ GeV, one finds
\begin{eqnarray}
Y+\ln Y=405.88,
\end{eqnarray}
which leads to the following value of the spectral index $n_s\cong 0.9700$, and a ratio of tensor to scalar perturbations equal to $r=0.1200$.

Further,  using \eqref{temperature} one obtains the following mass $m\sim 10^{-4} M_{pl}$. The value of the Hubble parameter at the transition time will be
$H_E\sim  10^{-6} M_{pl}$, and its value when the pivot scale leaves de Hubble radius is $H\sim   2\times 10^{-5} M_{pl}$. Finally the number of e-folds
is $N=66 $.

\item $T_R=10^9$ GeV, one finds
\begin{eqnarray}
Y+\ln Y=382.85,
\end{eqnarray}
which leads to the following value of the spectral index $n_s\cong 0.9682$, and a ratio of tensor to scalar perturbations equal to $r=0.1272$.

Further, using \eqref{temperature} one obtains the following mass $m\sim 10^{-9} M_{pl}$. The value of the Hubble parameter at the transition time will be
$H_E\sim 2\times 10^{-6} M_{pl}$, and its value when the pivot scale leaves de Hubble radius is $H\sim 4\times 10^{-5} M_{pl}$. Finally the number of e-folds
is $N=62$.

%Note that these quantities are incompatible with our assumption $m\gg H_E$ meaning that our model does not support this high reheating temperature.

\end{enumerate}

\
In fact, in Figure $1$ we show the corresponding ratio of tensor to scalar perturbations $r$ and mass $m$ obtained from a range of temperatures between $1$ MeV and $10^9$ GeV.
From
 these results we can conclude that:

\begin{itemize}
\item The theoretical value of the spectral index provided by the model ranges from $0.9682$ to $0.9723$
and, since Planck2015 data obtained the observational value $n_s=0.968\pm 0.006$, all the theoretical values of $n_s$ enter in the $1$-dimensional marginalized 
 2$\sigma$ C.L.
 
 \item Since $Y(T_R)$ is a decreasing function with the reheating temperature, and $r(T_R)=\frac{48}{Y(T_R)}$, one concludes that the tensor/scalar ratio is an
 increasing function with the reheating temperature. Therefore, for temperatures greater than $10^4$ GeV, our model does not support the observational data 
 given by the joint analysis of BICEP2/ Keck Array and Planck Data, where the $B$-mode polarization constrains the ratio of tensor to scalar perturbations to $r<0.12$
 at $2\sigma$ C.L. \cite{bicep2}. Moreover, since our formula \eqref{temperature} only holds for masses smaller than the Planck's one, one can conclude that, when the reheating is due to the 
 production of heavy massive particles, our model only supports reheating temperatures between $1$ GeV and $10^4$ GeV.
 \end{itemize}

\

\begin{figure}[H]
\begin{center}
\includegraphics[height=50mm]{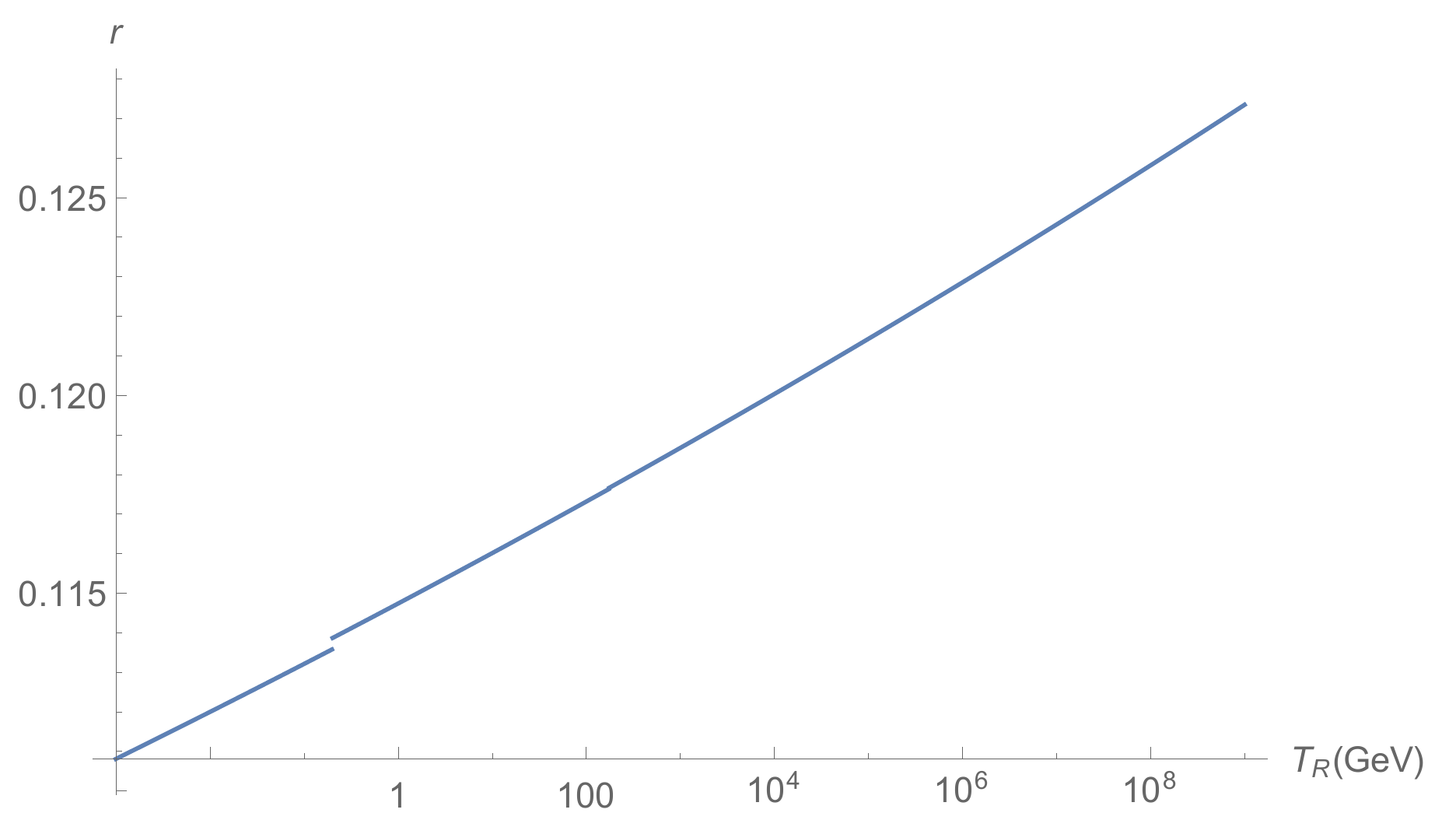}
\includegraphics[height=50mm]{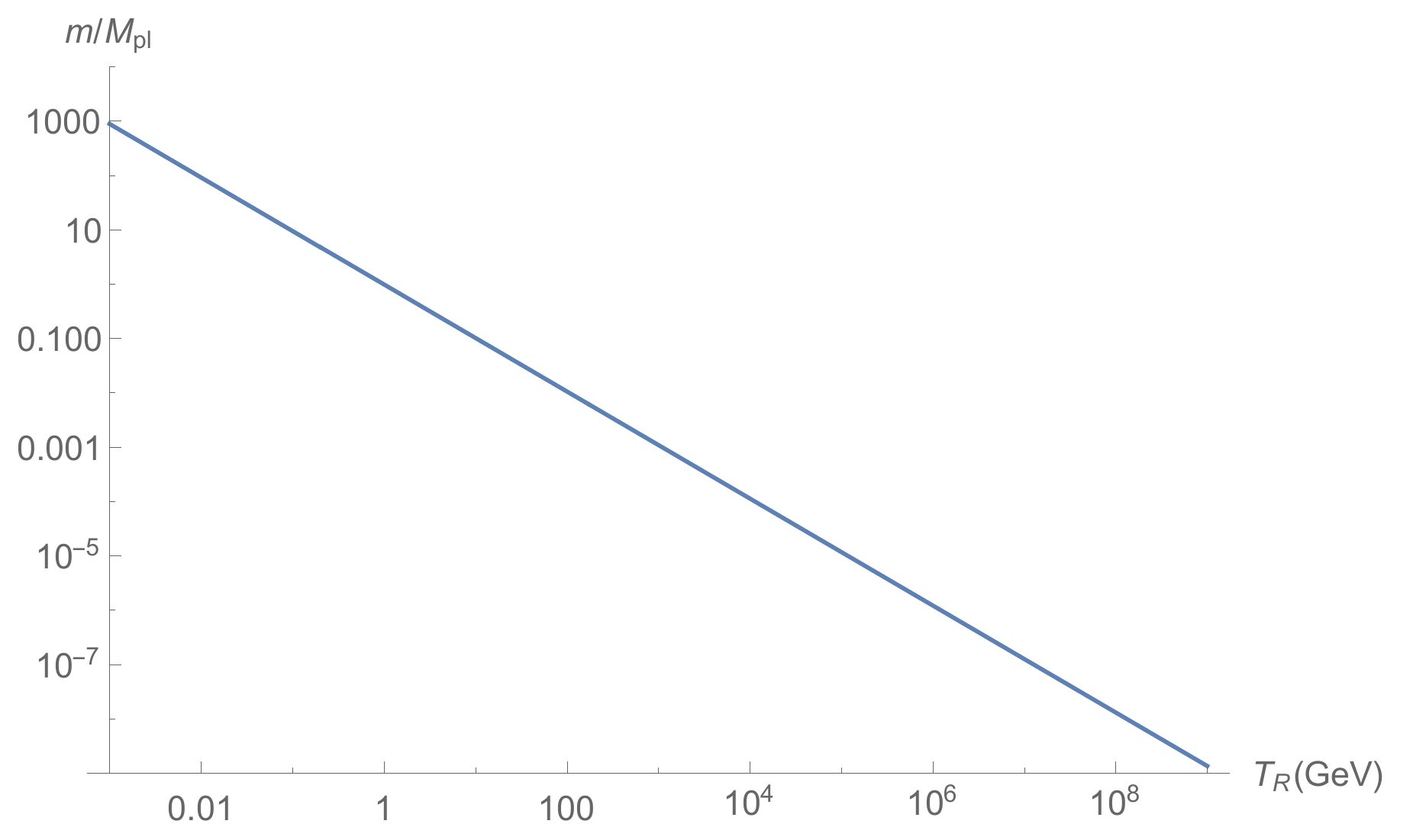}
\end{center}
\caption{Evolution of the tensor/scalar ratio $r$ (left) and of the mass of the produced particles $m$ (right) versus the  reheating temperature $T_R$.}
\end{figure}

 \subsection{Massless particles non-conformally coupled with gravity}

 As we have seen above using equation \eqref{X}, for our model the temperatures that lead to a ratio of tensor to scalar perturbations less than $0.12$ range between $1$ MeV and $10^4$ GeV. Therefore,
 using formula \eqref{temperature1} it is not difficult to check that, for these temperatures, the coupling constant $\xi$ must satisfy
 \begin{eqnarray}
  3\times 10^{-7}\lesssim \left|\xi-\frac{1}{6} \right| \lesssim  10^{-2}.
   \end{eqnarray}

 \
 
 On the other hand, dealing with massless particles, which are far from the conformal coupling with gravity, by using a toy model where there is a phase transition from de Sitter phase to radiation regime, it has been shown that the energy density of these produced particles is approximately \cite{Damour-Vilenkin}
 {\begin{eqnarray}
\rho_{\chi}\sim 10^{-2} H_E^4\left( \frac{a_E}{a} \right)^4,
\end{eqnarray}}
 which leads, after equating it to the reheating time with the energy density of the background, to the following reheating temperature
 \begin{eqnarray}
 T_R\sim 3\times 10^{-2}\frac{H_E^2}{M_{pl}^2}M_{pl}\sim 10^{-8}\left(\frac{1-n_s}{12} \right)^2 M_{pl},
 \end{eqnarray}
which together with equation \eqref{X} defines a system of two equations where the spectral index and the reheating temperature are the unknown variables. 

\

Unfortunately, the system is only compatible, i.e. has solutions, for temperatures greater than $10^5$ GeV, leading to a tensor/scalar ratio greater than $0.12$, which disagrees with
the recent observational data \cite{bicep2} meaning that the model does not work for particles with a conformal coupling far from $1/6$. Consequently, we can conclude that our model only supports the production of massless particles with a coupling constant very close to $1/6$, i.e. nearly conformally coupled with gravity, leading to a reheating temperature ranging between
$1$ MeV and $10^4$ GeV.

 \section{Conclusions}
 In the  present work we have  obtained formula \eqref{X}, which relates the spectral index with the reheating temperature. Since for the quintessential inflation model studied here the
spectral index ($n_s$) and the ratio of tensor to scalar perturbations $(r)$ are related through $r= 4(1-n_s)$, from the observational constraint $r\leq 0.12$ \cite{bicep2} one can deduce using \eqref{X} that the reheating temperature must be below $10^4$ GeV. 

\
 
On the other hand, the simplest way to reheat the universe is via the production, at the phase transition from inflation to kination, of very heavy massive particles conformally coupled with gravity or massless particles. In the former case, since elementary particles with masses around the Planck's mass become micro black holes, whose physics is unknown, one has to consider the production of particles with mass smaller than the Planck's one, which leads to a reheating temperature greater than $1$ GeV. { Moreover, the constraint for the mass of the produced particles, $10^{-4}M_{pl}\leq m\leq M_{pl}$, has also been found by using expression \eqref{rhomassive}, which refers to the energy density of massive produced particles as is extensively proved in the Appendix.}

\

Finally, dealing with the creation of massless particles, we have shown that the viability of the model is only possible when these particles are nearly conformally coupled with gravity, because when the coupling constant is far from $1/6$ one obtains reheating temperatures greater than $10^5$ GeV, which is incompatible with the fact that the tensor/scalar ratio must be smaller than $0.12$.

 \section{Appendix}

  In this appendix we will calculate the one-loop energy density  due to  a heavy massive quantum field, namely $\chi$,  conformally coupled with gravity.
 This quantity is given by \cite{Birrell}
 \begin{eqnarray}
 \rho_{\chi}(\tau)=\frac{1}{4\pi^2 a^4(\tau)}\int_0^{\infty}(|\chi_k'|^2+(k^2+m^2a^2(\tau))|\chi_k |^2)k^2dk,
 \end{eqnarray}
  where $\chi_k$ is the $k$-mode associated to the vacuum. If initially, at some initial time namely $\tau_i$, the quantum field is in the vacuum state, then the modes
  must satisfy at that time 
  \begin{eqnarray}
  \chi_k(\tau_i)=\frac{e^{-i\int^{\tau_i}\omega_k(\tau)d\tau}}{\sqrt{2\omega_k(\tau_i)}}, \quad 
  \chi_k'(\tau_i)=-i\sqrt{\frac{\omega_k(\tau_i)}{2}} {e^{-i\int^{\tau_i}\omega_k(\tau)d\tau}}, \end{eqnarray} 
  where $\omega_k(\tau)=\sqrt{k^2+m^2a^2(\tau)}$ is the frequency of the $k$-mode.
  
  Inserting this quantity in the one-loop energy density one obtains
 \begin{eqnarray}
 \rho_{\chi}(\tau_i)=\frac{1}{4\pi^2 a^4(\tau_i)}\int_0^{\infty} \omega_k(\tau_i)k^2dk,
 \end{eqnarray}
 which is the zero-point energy density of the vacuum. This divergent quantity could be removed from different ways: the simplest one is to subtract to the one loop energy density the zero-point energy density, obtaining the following convergent quantity (see \cite{gmm})
 \begin{eqnarray}
 \rho_{\chi}^{conv}(\tau)=\frac{1}{4\pi^2 a^4(\tau)}\int_0^{\infty}(|\chi_k'|^2+(k^2+m^2a^2(\tau))|\chi_k |^2- \omega_k(\tau))k^2dk. \label{rho0pre}
 \end{eqnarray} 

In order to compute this integral, we use that the $\chi_k$ mode can be approximated by the WKB solution of 2n-th order, namely $\chi_{k,WKB}(\tau)=\frac{1}{\sqrt{W_{2n,k}(\tau)}}e^{-i\int^{\tau}W_{2n,k}(\eta)d\eta}$, defined by the recurrence \cite{Haro}
\begin{eqnarray}
 W_{2n,k}(\tau)=\left\{ \begin{array}{ll}  \omega_k(\tau), & \mbox{$n=0$} \\ \omega_k(\tau)-\frac{1}{2\omega_k}\left[\frac{W_{2(n-1),k}''(\tau)}{2W_{2(n-1),k}(\tau)}-\frac{3}{4}\left(\frac{W_{2(n-1),k}'(\tau)}{W_{2(n-1),k}(\tau)} \right)^2 \right], & \mbox{$n\in\mathbb{N}$.} \end{array} \right.
\end{eqnarray}

Then, the integral in \eqref{rho0pre} in order zero becomes $\rho_{\chi}^{conv}\approx\frac{m^2H^2}{96\pi}$. However, this prescription only holds for conformally coupled fields. When one deals with non-conformally coupled fields one of the most populars ways to renormalize the energy density is to use the adiabatic regularization, which consists in subtracting the zero, second and fourth order adiabatic expressions of the energy density (see for instance \cite{Bunch}). In this way, as has been showed in \cite{kaya}, when $m\gg H$ the renormalized energy density, namely $\rho^{ren}_{\chi}(\tau)$ is of the order $\frac{H^6}{m^2}$, then for values of the Hubble parameter satisfying $H\leq M_{pl}$ and $m\gg H$,  since 
$\frac{H^6}{m^2 }\ll H^2 M_{pl}^2$ one can conclude that back-reaction does not affect the dynamics of the background. 

\

This condition is broken at the phase transition, because there the WKB approximation cannot hold and the negative and positive frequencies mix. Hence, the $\chi_k$  mode becomes  
%$\mu_k:=a_k\chi_{k}+ b_k\chi_{k}^*$ that has 
approximately of the form $\alpha_k\chi_{k,WKB}+ \beta_k\chi_{k,WKB}^*$, where the $\beta$-Bogoliubov coefficient is given by $\beta_k=-i\mathcal{W}[\chi_{k}(\eta_E^-),\chi_{k}(\eta_E^+)]$ and verifying that $|\alpha_k|^2-|\beta_k|^2=1$. Thus, the energy density becomes 
%the following one by replacing $\mu_k$ by $\chi_k$ in equation \eqref{rho0pre}:
\begin{eqnarray}
\rho_{\chi}^{ren}(\tau)=\frac{1}{4\pi^2 a^4(\tau)}\int_0^{\infty}\left\{\left[|\chi_{k.WKB}'|^2+\omega_k^2(\tau)|\chi_{k,WKB} |^2\right](1+2|\beta_k|^2)\right.\nonumber\\ \left.+2\mathcal{R}e(\alpha_k\beta_k^*\left[(\chi_{k,WKB}')^2+\omega_k^2(\tau)(\chi_{k,WKB})^2\right])\right\}k^2dk-\rho_{4}^{ adia},
\end{eqnarray}
where  $\rho_{4}^{adia}$ contains all the adiabatic terms up to order 4.

Since, as has been showed in \cite{kaya}, the term
\begin{eqnarray}
\frac{1}{4\pi^2 a^4(\tau)}\int_0^{\infty}\left[|\chi_{k.WKB}'|^2+\omega_k^2(\tau)|\chi_{k,WKB} |^2\right]k^2dk-\rho_{4}^{ adia},\end{eqnarray}
is of the order $H^6/m^2$,
using that $\alpha_k\simeq 1$ and the approximation to zero order $\chi_{k,WKB}(\eta)\approx\frac{e^{-i\int^{\eta}\omega_k(\tau)d\tau}}{\sqrt{2\omega_k(\eta)}}$, we obtain
\begin{eqnarray}
\rho_{\chi}^{ren}(\tau)\approx \frac{1}{4\pi^2a^4(\tau)}\int_0^{\infty}\left[\frac{(\omega_k'(\tau))^2}{4\omega_k^3(\tau)}|\beta_k|+2\omega_k(\tau)|\beta_k|^2\right]k^2dk.
\end{eqnarray}

In \cite{hap2}, it was already computed that $|\beta_k|\approx \frac{9m^2a_E^5H_E^3}{8(k^2+m^2a_E^2)^{5/2}}$. Therefore, we obtain 
\begin{eqnarray}
\rho_{\chi}^{ren}\approx \frac{81H_E^6}{128\pi^2m^2}\left(\frac{a_E}{a} \right)^4\int_0^{\infty}\frac{x^2\sqrt{x^2+\left(\frac{a}{a_E}\right)^2}}{(x^2+1)^5}dx+\frac{9H^2H_E^3}{128\pi^2m}\left(\frac{a}{a_E}\right)^2\int_0^{\infty}\frac{x^2dx}{\left(x^2+\left(\frac{a}{a_E}\right)^2\right)^{5/2}(x^2+1)^{5/2}}.
\end{eqnarray}

\

First of all we analyze the second term, which can be bounded by
\begin{eqnarray}\label{xxx}
\frac{9H^2H_E^3}{128\pi^2m}\left(\frac{a_E}{a}\right)^3\int_0^{\infty}\frac{x^2dx}{(x^2+1)^{2}}=
\frac{9H^2H_E^3}{512\pi m}\left(\frac{a_E}{a}\right)^3.
\end{eqnarray}
Since the energy density of the background evolves as $H^2M_{pl}^2$ 
and  $H_E<m<M_{pl}$ one can easily check that \eqref{xxx} is subdominant with respect to the background.

\

To deal with the first term, we note that after the phase transition the universe enters into a kination regime, whose dynamics are given by $\dot{H}\cong -3H^2$. Then, with a simple calculation, we find that the time needed after the phase transition to have $a=10^2a_E$ is of the order $\frac{10^6}{H_E}\sim 10^{-32}s$. 
This is a very small time compared with the time after the phase transition to reheat the universe, which is of the order $10^{-10} s$ \cite{haro2016}. Then, one can assume that
from the phase transition to the end of the reheating process $a/a_E\gg 1$, and following the calculations performed in \cite{he} one can make the approximation
$\sqrt{x^2+\left(\frac{a}{a_E}\right)^2}\cong \frac{a}{a_E}$, to finally obtain
\begin{eqnarray}
\rho_{\chi}^{ren}\approx \frac{405H_E^6}{32768\pi m^2}\left(\frac{a_E}{a}\right)^3\sim 4\times 10^{-3}\frac{H_E^6}{m^2}\left(\frac{a_E}{a}\right)^3.
\end{eqnarray}

 % \newpage

\

%\vspace{0.5cm}
\section*{Acknowledgments} We would like to thank Professor Jaume Amor\'os for carefully reading the manuscript.
This investigation has been supported in part by MINECO (Spain), project MTM2014-52402-C3-1-P.
%and MTM2012-38122-C03-01. 

\end{document}